\documentclass[aps,prd,12point,twocolumn,nofootinbib,showpacs,superscriptaddress]{revtex4-1}

\usepackage{graphicx}
\usepackage{dcolumn}
\usepackage{tabu}
\usepackage{amsmath,amssymb}
\usepackage{float}
\usepackage{natbib}
\usepackage{balance}
\usepackage{braket}
\usepackage{mathrsfs}
\usepackage{bm}
\usepackage{lipsum}
\usepackage[dvipsnames]{xcolor}
\usepackage[title]{appendix}
\usepackage[breaklinks=true,pdfborder={0 0 0},bookmarksopen]{hyperref}
\hypersetup{
	colorlinks   = true, 
	urlcolor     = magenta, 
	linkcolor    = blue, 
	citecolor   = magenta 
}

\begin{document}

\title{Noether charge and black hole entropy in teleparallel gravity}


\author{F. Hammad} \email{fhammad@ubishops.ca}
\affiliation{Department of Physics and Astronomy, Bishop's University, 2600 College Street, Sherbrooke, QC, J1M~1Z7 Canada}
\affiliation{Physics Department, Champlain 
College-Lennoxville, 2580 College Street, Sherbrooke,  
QC, J1M~0C8 Canada}
\affiliation{D\'epartement de Physique, Universit\'e de Montr\'eal,\\
2900 Boulevard \'Edouard-Montpetit,
Montr\'eal, QC, H3T 1J4
Canada}
\author{D. Dijamco} \email{ddijamco18@ubishops.ca} 
\affiliation{Department of Physics and Astronomy, Bishop's University, 2600 College Street, Sherbrooke, QC, J1M~1Z7 Canada}
\author{A. Torres-Rivas} \email{andoni.torres-rivas@usherbrooke.ca}
\affiliation{D\'epartement de Physique, Universit\'e de Sherbrooke, Sherbrooke, QC, J1K~2X9 Canada} 
\affiliation{Department of Physics and Astronomy, Bishop's University, 2600 College Street, Sherbrooke, QC, J1M~1Z7 Canada}
\author{D. B\'erub\'e} \email{damien.berube.qc@gmail.com} 
\affiliation{Physics Department, Champlain 
College-Lennoxville, 2580 College Street, Sherbrooke, QC, J1M~0C8 Canada}

\begin{abstract}
The Noether charge associated to diffeomorphism invariance in teleparallel gravity is derived. It is shown that the latter yields the ADM mass of an asymptotically flat spacetime. The black hole entropy is then investigated based on Wald's prescription that relies on the Noether charge. It is shown that, like in general relativity, the surface gravity can be factored out from such a charge. Consequently, the similarity with the first law of thermodynamics implied by such an approach in general relativity does show up also in teleparallel gravity. It is found that, based on the expression of the first law of black hole mechanics, which is preserved in teleparallel gravity, entropy can thus be extracted from such a Noether charge. The resulting entropy can very naturally be expressed as a {\it volume integral}, though. As such, it is shown that the conformal issue that plagues the entropy-area law within general relativity does not arise in teleparallel gravity based on Wald's approach. The physics behind these features is discussed.
\end{abstract}

\pacs {04.20.Cv, 04.50.Kd, 04.70.-s}
\maketitle
\section{Introduction}\label{sec:Intro}
Two of the major motivations that are often put forward in the search for gravitational theories beyond general relativity (GR) are the quantization problem and the dark energy and/or dark matter problems. The usual cure suggested to these problems consists in modifying the Einstein-Hilbert Lagrangian (see e.g., the very recent review in Ref.~\cite{Motivation}). Thus, nonlinear functionals of the Ricci scalar $R$ and other geometric invariants, as well as extra degrees of freedom for spacetime, such as torsion and scalar fields, are often added inside the simple GR Lagrangian. However, while these extra terms and entities do indeed enrich the theory, they do make the gravitational action, let alone the field equations, exceedingly complex. In addition, any specific extra term added to the GR Lagrangian automatically begs for a justification from first principles. In this regard, the only theory that is equivalent to GR, and yet simpler and richer than GR, is arguably the so-called teleparallel equivalent to general relativity (TEGR) (see Ref.~\cite{TEGR} for a textbook introduction and Ref.~\cite{MalufReview} for a review). 

TEGR is known to incorporate the nice features of GR, such as the possibility of studying conserved currents \cite{ConservedCurrents} and nonvacuum solutions \cite{NonVacuumSolutions}. Furthermore, TEGR has been shown to be an adequate framework for studying gravitational waves \cite{TEGRGW1,TEGRGW2}, nonsingular black holes \cite{NonsingularBHinTEGR} and energy fluxes in cylindrical spacetimes \cite{CylinderTEGR}. TEGR also makes it possible to come up with extensions of GR to tackle the problems of dark matter \cite{NonlocalViaTEGR1,NonlocalViaTEGR2} and cosmology \cite{Cosmology}.

The feature of TEGR that makes it simpler than GR is its first order Lagrangian as well as its Yang-Mills-like field equations. This feature arises thanks to the use of the frame fields which can be thought of as representing the ``square root'' of the metric. The main feature of the TEGR framework that makes it richer than that of GR \cite {TEGR=GR?} is the fact that, unlike in GR, one is able to separate gravitational effects from inertial effects. This is achieved thanks to the spin connection that is contained inside the Weitzenb\"ock connection. This is a unique feature to teleparallel theories of gravity in general \cite{NewGR,Motivation}, and TEGR in particular, that is not shared by any other alternative theories of gravity. This feature of the theory is behind the remarkable possibility of defining an energy-momentum tensor for the gravitational field \cite{MalufEnergy,MalufPressure}. It is thus of great interest to investigate within this theory the fate of black hole thermodynamics as well. In fact, the problem of black hole entropy with its peculiar area-law character is of tremendous importance as it depends --- in contrast to temperature --- on the dynamics of spacetime rather than on the kinematics of the latter. Thereby, with its capacity for splitting the dynamics into inertial effects plus purely gravitational effects, TEGR constitutes probably the best framework in which one might still hope to learn more about black hole entropy at the classical level.

It has recently been shown in Ref.~\cite{AugustPaper} that the analogy between the first law of black hole mechanics and thermodynamics becomes spoiled under a Weyl conformal transformation of spacetime. One of the reasons is that the area-law character of black hole entropy becomes problematic. It was found indeed that, in contrast to a surface area, the black hole entropy based on Wald's approach remains invariant under a conformal transformation. In Ref.~\cite{FebruaryPaper}, it was shown in this regard that even the extraction of Einstein's field equations based on Jacobson's prescription \cite{JacobsonPRL} becomes problematic. A first clue of the general issue is actually provided by the observation that under conformal transformations of spacetime a black hole horizon might disappear altogether \cite{Hammad1} and a wormhole, which requires the violation of the null energy condition (NEC) by matter, becomes sustained without violating the NEC as soon as one transforms spacetime {\it\`a la} Weyl \cite{OctoberPaper}. The very root of the problem can therefore be traced back to the fundamental dichotomy between matter and geometry that is inherent in Einstein's field equations. As matter and geometry behave differently under a conformal transformation \cite{Hammad2,Hammad3}, a Weyl transformation necessarily filters out the material entities from the geometric ones in any dynamical relation that involves both of them.   
 
In light of this first clue revealed by the Weyl transformation, our goal in this paper is to use TEGR to refine this picture by investigating how much the intricate relation between inertia and geometry is responsible in shaping black hole entropy. In fact, on the one hand, we have the simple dichotomy between matter and geometry that betrays the fundamental nature of the entropy area law when examined through the lens of Weyl transformations. On the other hand, inertia, being an intimate property of pure matter, becomes ``unfortunately" geometrized in GR along with pure gravity. It is therefore very tempting to believe that within a framework in which gravitational and inertial effects could be made distinct, another picture of black hole entropy could emerge. 

Indeed, as mentioned above, TEGR is already known to offer, in contrast to GR, a framework in which the energy-momentum density for the gravitational field is well defined. This is made possible in TEGR thanks to the separability of the energy-momentum density's {\it pseudotensor} of inertial effects from the purely {\it tensorial} gravitational contribution \cite{TEGR}. Remarkably, the Arnowitt-Deser-Misner (ADM) mass obtained from such an energy-momentum tensor of the gravitational field is more naturally expressed as a volume integral \cite{MalufEnergy,MalufPressure}, in contrast to what is found within GR where it takes the form of a boundary integral. The question that arises then is: Could it be possible that in TEGR entropy also becomes more naturally encoded in the volume of space inside the black hole horizon like the ordinary thermodynamic concept of entropy? This question is intimately linked with the well-known fact that an observer in an accelerating frame automatically witnesses an entropy that also obeys the area law. In agreement with the strong equivalence principle, the observer indeed ``feels" locally the same force as that of real gravity but only thanks to the purely inertial effects. As such, one might therefore wonder whether the familiar area law of black hole entropy in GR is not actually simply due to the mixing between real gravitational effects and purely inertial effects. In other words, is it possible that in a theory, like TEGR, in which inertia can be separated from gravity, black hole entropy could also be ``purified" from inertial effects and allowed to reveal its true nature? Our aim in this paper is to investigate such a question in detail. Furthermore, to the best of our knowledge the Noether charge approach has not been previously applied to TEGR.

The following sections of the paper are organized as follows. In Sec.~\ref{sec:Charge}, we introduce the main definitions and equations of TEGR. We then use these to extract the symplectic potential from which the Noether charge associated with the diffeomorphism invariance of the theory can be derived according to Wald's prescription. In Sec.~\ref{sec:ADM}, we show that such a Noether charge provides the gravitational energy of an asymptotically flat spacetime which is just the familiar ADM mass derived from GR. In Sec.~\ref{sec:Entropy}, we use the Noether charge to extract the black hole entropy. In Sec.~\ref{sec:UnderWeyl}, we examine the behavior of such an entropy under Weyl transformations. We conclude this paper with a brief summary and discussion section.
\section{Noether charge in TEGR}\label{sec:Charge}
Like all teleparallel gravity theories, TEGR is a diffeomorphism-invariant theory. Therefore, to extract the Noether charge associated to its diffeomorphism invariance we apply Wald's algorithm \cite{WaldEntropy,IyerWald}\footnote{Wald's ``algorithm" is actually more than just an algorithm. It is so fundamental that it allows one to apply it to extended theories beyond GR \cite{JacobsonKangMyers}, and even allows one to search for new modified gravity theories \cite{HammadfR}.}. The first step of the algorithm is to write down the Lagrangian of the theory in the language of differential forms. 

The dynamical field in TEGR is the frame (or tetrad, or vierbein) field $e^a_\mu(x)$, defined through its relation to the metric $g_{\mu\nu}(x)$ of curved spacetime by, $g_{\mu\nu}=\eta_{ab}e^a_\mu e^b_\nu$, where $\eta_{ab}$ is the flat Minkowski metric\footnote{Throughout the paper, we denote, as customary, the flat tangent-space (Lorentz) indices by Latin letters while we reserve the Greek letters to denote the curved-spacetime indices.}. The inverse vierbein fields are denoted by $e^\mu_a$ such that, $e_a^\mu e^a_\nu=\delta^\mu_\nu$ and $e_a^\mu e^b_\mu=\delta^b_a$. Then, the metric determinant $g$ is expressed as $\sqrt{-g}=\det(e^a_\mu)\equiv e$. The affine connection of TEGR is the Weitzenb\"ock connection $\Gamma^\mu_{\nu\rho}$, given in terms of the frame fields and the spin connection $\omega_\mu\!\,^a\!\,_b(x)$ by,
\begin{equation}\label{AffineConnection}
    \Gamma^\mu_{\nu\rho}=e^\mu_a\partial_\rho e^a_\nu+e^\mu_a \,\omega_\rho\!\,^a\!\,_b\,e^b_\nu.
\end{equation}
In contrast to Einstein-Cartan gravity, the spin connection $\omega_\mu\!\,^a\!\,_b(x)$ in TEGR is chosen to be a purely inertial Lorentz connection. That is, it is built pointwise from a local Lorentz transformation $\Lambda^a\!\,_b(x)$, so that  $\omega_\mu\!\,^a\!\,_b(x)=(\Lambda^{-1})^a\!\,_c\,\partial_\mu\Lambda^c\!\,_b$. This specific structure of the affine connection is indeed what allows one to separate in teleparallel gravity inertial effects from purely gravitational effects. In fact, since under a local Lorentz transformation the spin connection transforms as $\omega'_\mu\!\,^a\!\,_b=(\Lambda^{-1})^a\!\,_c\,\omega_\mu\!\,^c\!\,_d\,\Lambda^d\!\,_b+(\Lambda^{-1})^a\!\,_c\,\partial_\mu\Lambda^c\!\,_b$, one can always start from a globally vanishing spin connection and perform a local Lorentz transformation to arrive at such a spin connection. Conversely, given any spin connection, one can perform local Lorentz boosts so that the final spin connection vanishes globally, canceling thus the purely inertial effects. This would then leave in the affine connection (\ref{AffineConnection}) only the tetrad fields' contribution which is thus due to pure gravity. In this case, the Weitzenb\"ock connection (\ref{AffineConnection}) reduces to $\Gamma^\mu_{\nu\rho}=e^\mu_a\partial_\rho e^a_\nu$. This specific form of the connection gives rise to the absolute parallelism condition which is often imposed and thus assumed in TEGR. 

From the Weitzenb\"ock connection one builds the torsion tensor, $T^\mu\,\!_{\nu\rho}:=\Gamma^\mu_{\rho\nu}-\Gamma^\mu_{\nu\rho}$, and from the latter one constructs the contortion tensor, $K^\mu\,\!_{\nu\rho}$, given, respectively, by,
\begin{align}\label{Torsion+Contortion}
    T^\mu\!\,_{\nu\rho}&=e^\mu_a\left(\partial_\nu e^a_\rho-\partial_\rho e^a_\nu+ \omega_\nu\!\,^a\!\,_b\,e^b_\rho-\omega_\rho\!\,^a\!\,_b\,e^b_\nu\right),\nonumber\\
    K^\mu\,_{\nu\rho}&=\tfrac{1}{2}\left(T_\nu\,\!^\mu\,\!_\rho+T_\rho\,\!^\mu\,\!_\nu-T^\mu\,\!_{\nu\rho}\right).
\end{align}
From these two entities one builds the so-called superpotential tensor, $\,\mathcal{S}_\rho\,\!^{\mu\nu}={K^{\mu\nu}}_\rho-\delta^\nu_\rho\, T^\mu+\delta^\mu_\rho\, T^\nu$, which is antisymmetric in its last two indices. The trace of torsion is here defined by $T^\mu={T^{\sigma\mu}}_\sigma$. With these ingredients, one finally introduces the contraction $\mathbb{T}={\mathcal{S}_\rho}^{\mu\nu}{T^\rho}_{\mu\nu}$ from which the action of TEGR is built:
\begin{align}\label{TPAction}
    \frac{1}{32\pi}\int {\rm d}^4x\,e\,\mathbb{ T}&=\frac{1}{32\pi}\int \mathbb{T}\boldsymbol{\epsilon}.
\end{align}
Here, we have introduced the Levi-Civita tensor $\boldsymbol{\epsilon}_{\mu\nu\rho\sigma}=e\,\epsilon_{\mu\nu\rho\sigma}$, where $\epsilon_{\mu\nu\rho\sigma}$ is the totally antisymmetric Levi-Civita symbol, defining thus a volume 4-form. Thus, in the language of forms the Lagrangian of TEGR reads, ${\bf L}=\mathbb{T}\boldsymbol{\epsilon}/32\pi.$

Next, Wald's algorithm for extracting the Noether charge consists in performing the following steps. First, vary the Lagrangian with respect to the dynamical fields of the theory. In TEGR these are the tetrad fields. So we get $\delta{\bf L}={\bf E}_a\,\!^\mu\,\delta e^a_\mu+{\rm d}{\boldsymbol\Theta}$, where
${\bf E}_a\,\!^\mu$ stands for the equations of motion 4-form (see Appendix \ref{A}). The so-called symplectic potential $\boldsymbol{\Theta}$ is a 3-form, and is here found to be given by (see Appendix \ref{B}),
\begin{equation}\label{Symplectic}
    \boldsymbol{\Theta}_{\beta\gamma\lambda}=-\frac{1}{8\pi}\delta e_\mu^a{\mathcal{S}_a}^{\mu\alpha}\boldsymbol{\epsilon}_{\alpha\beta\gamma\lambda}.
\end{equation}
From this symplectic potential, one extracts the Noether charge associated to the diffeomorphism generated by a vector field $\xi^\mu$ by, first, replacing the general variation $\delta e^a_\mu$ in Eq.~(\ref{Symplectic}) by the Lie derivative $\pounds_\xi e^a_\mu$ induced by the diffeomorphism. Then, one notes that when the equations of motion are satisfied, {\it i.e.}, when ${\bf E}_a\,\!^\mu=0$, the Noether current 3-form, ${\bf J}=\boldsymbol{\Theta}-i_\xi{\bf L}$ (where $i_\xi{\bf L}$ stands for the interior derivative of $\bf L$ with respect to the vector $\xi^\mu$) is a closed form, {\it i.e.}, ${\rm d}{\bf J}=0$. For this means indeed that there exists the charge 2-form $\bf Q$, such that ${\bf J}={\rm d}{\bf Q}$ (see the explicit derivation in Eq.~(\ref{dJ})). This charge 2-form is the so-called Noether charge \cite{WaldEntropy,IyerWald}. 

However, from the expression (\ref{Symplectic}) of the symplectic potential, it seems {\it a priori} straightforward that, when $\xi^\mu$ is taken to be the Killing vector field of the spacetime, as is done to study black hole thermodynamics, one automatically has $\pounds_\xi e^a_\mu=0$, leading to an identically vanishing symplectic potential. Within the framework of TEGR, though, things become much less trivial. In fact, one has there, in addition to pure gravity, also inertial effects built-in on a separate footing. These manifest themselves by the fact that any infinitesimal displacement in spacetime is automatically accompanied by a Lorentz rotation of the frame fields. Thus, the full Lie derivative of the frame fields should read instead\footnote{This is actually another independent case for the need to introduce a generalized Lie derivative besides the motivations for a similar derivative operator given in Refs.~\cite{ObukhovRubilar,LorentzLie1,LorentzLie2}.}, $\mathcal{L}_\xi e^a_\mu:=\pounds_\xi e^a_\mu+{\mathcal{M}^a}_b(x,\xi) e^b_\mu$. Here, the term $\pounds_\xi e^a_\mu$ stands for the usual Lie derivative of a covariant spacetime vector, $\pounds_\xi e^a_\mu=\xi^\nu\partial_\nu e^a_\mu+e^a_\nu\partial_\mu\xi^\nu$, whereas $\mathcal{M}^a\,\!_b(x,\xi)$ denotes an antisymmetric Lorentz rotation matrix. The latter acts on the Lorentz index of the tetrad and should {\it a priori} depend on spacetime as well as on the Killing vector field $\xi^\mu$. With such a generalized operator, one might indeed impose that $\mathcal{L}_\xi e^a_\mu=0$, and yet $\pounds_\xi e^a_\mu\neq0$. 

Notwithstanding this feature of TEGR, given that in the latter one can always switch to a class of reference frames in which the rotation matrix $\mathcal{M}^a\,\!_b(x,\xi)$ vanishes identically, the Killing condition which emerges from $\pounds_\xi g_{\mu\nu}=0$ can thus always be reduced again to $\pounds_\xi e^a_\mu=0$. For the sake of generality, however, we have derived in Appendix \ref{B} the Noether charge using such a generalized Lie derivative. We thus came to the conclusion there that the Lorentz rotation matrix $\mathcal{M}^a\,\!_b$ has to satisfy a very specific condition to guarantee the existence of the symplectic potential $\boldsymbol\Theta$. 

Actually, thanks to the possibility of choosing arbitrary Lorentz reference frames in TEGR without changing the dynamics of the theory, one can still satisfy the Killing condition $\pounds_\xi g_{\mu\nu}=0$ with the weaker requirement on the tetrad fields to satisfy, $e_{a\nu}\pounds_\xi e^a_\mu=-e_{a\mu}\pounds_\xi e^a_\nu$, rather than $\pounds_\xi e^a_\mu=0$. However, as it will be discussed in Sec.~\ref{sec:ADM}, the first law of black hole mechanics within Wald's approach requires one to have $\pounds_\xi \phi=0$ satisfied by all the dynamical fields $\phi$ of the theory. These dynamical fields just happen to be solely the tetrad fields in TEGR. 

Following now Wald's algorithm, for extracting the Noether charge by combining the symplectic potential (\ref{Symplectic}) and the Lagrangian in Eq.~(\ref{TPAction}), we easily find the following charge 2-form (see Appendix \ref{B}),
\begin{equation}\label{Charge}
{\bf Q}_{\gamma\lambda}[\xi]=\frac{1}{16\pi}\xi^a{\mathcal{S}_a}^{\alpha\beta}\boldsymbol{\epsilon}_{\alpha\beta\gamma\lambda}.
\end{equation}
This charge is supposed, according to Wald's approach, to yield the entropy of a black hole when the diffeomorphism generator $\xi^\mu$ is taken to be the Killing vector field of the spacetime and when such a charge is integrated over the bifurcation 2-surface of the horizon. Before we examine that in detail, however, we need first to make sure that, as in GR \cite{IyerWald}, the integral over a closed spatial boundary at infinity of the variation of such a charge, when combined with the symplectic potential (\ref{Symplectic}), does yield the variation $\delta\mathcal{E}$ of the energy enclosed inside the boundary. For a timelike Killing vector and a boundary at spatial infinity such an energy should, as in GR, coincide with the ADM mass.
\section{ADM mass and Noether charge}\label{sec:ADM}
As shown in detail in Ref.~\cite{IyerWald}, if a 3-form $\bf B$ exists, such that $\int_\infty i_\xi\boldsymbol{\Theta}=\delta\int_\infty i_\xi \bf B$, then a Hamiltonian $H$ describing the dynamics generated by the vector $\xi^\mu$ does exist also and is given on shell by $H=\int_\infty({\bf Q}[\xi]-i_\xi\bf B)$. For an asymptotic time translation $t^\mu$, the canonical energy $\mathcal{E}$ can then be defined to be $\mathcal{E}=\int_\infty({\bf Q}[t]-i_t\bf B)$ \cite{IyerWald}. In the case of GR, it was shown that the 3-form $\bf B$ does indeed exist and that $\mathcal{E}$ coincides with the ADM mass \cite{IyerWald}. Our goal in this section is to check whether this remains true in TEGR when we use our previous expressions (\ref{Symplectic}) and (\ref{Charge}) of $\boldsymbol{\Theta}$ and $\bf{Q}$, respectively. 

For an asymptotically flat spacetime, we have $e^a_\mu=\delta^a_\mu+\mathcal{O}(1/r)$, where $\mathcal{O}(1/r)$ denotes terms that decrease like $1/r$ at infinity \cite{MalufEnergy,MalufPressure}. Therefore, given that the superpotential $\mathcal{S}_a\,\!^{\mu\nu}$ is built from the torsion tensor (\ref{Torsion+Contortion}), which is made of the spin connection and the first derivatives of the frame fields, we learn that at spatial infinity we have $\mathcal{S}_a\,\!^{\mu\nu}\sim\mathcal{O}(1/r^2)$. In fact, on the one hand, the spin connection of TEGR is proportional to the derivative $\partial_\mu\Lambda^a\,\!_b$ of the Lorentz matrix. On the other hand, the Lorentz matrix representing the inertial effects in TEGR should decrease at least as fast as $1/r$ in an asymptotically flat spacetime. Therefore, on the spatial boundary at infinity we have,
\begin{equation}\label{tTheta}
    \int_\infty \xi^\beta\boldsymbol{\Theta}_{\beta\gamma\lambda}=-\frac{1}{8\pi}\int_\infty \xi^\beta\delta e_\mu^a{\mathcal{S}_a}^{\mu\alpha}\boldsymbol{\epsilon}_{\alpha\beta\gamma\lambda}=0.
\end{equation}
Thereby, the Hamiltonian in this case is simply given by $H=\int_\infty{\bf Q}[\xi]$. Making use of the Noether charge 2-form (\ref{Charge}), we have then the following energy enclosed by a closed 2-sphere at spatial infinity on which the vector field $\xi^\mu$ is taken to be the asymptotic time translation:
\begin{align}\label{ADM}
\mathcal{E}&=\int_\infty{\bf Q}[\xi]\nonumber\\
&=\frac{1}{16\pi}\int_\infty t^a{\mathcal{S}_a}^{\alpha\beta}\boldsymbol{\epsilon}_{\alpha\beta\gamma\lambda}\nonumber\\
&=\frac{1}{8\pi}\int_{V}{\rm d}^3x\, \partial_i(e{\mathcal{S}_0}^{0i})\nonumber\\
&=M_{\small{ADM}}.
\end{align}
In the third line we have used Stokes' theorem to turn the surface integral into an integral over the volume $V$, and in the last line we used the fact that the ADM mass in TEGR is specifically given by such a volume integral \cite{MalufEnergy,MalufPressure}. Notice that in contrast to the ADM mass in GR, the formula we find here is more naturally expressed as a volume integral. This result allows us now to discuss the black hole entropy.  
\section{Black hole entropy}\label{sec:Entropy}
The first law of black hole mechanics for a static black hole of mass $M=\mathcal{E}$, as derived within GR, reads \cite{BCH,Wald}, $\delta\mathcal{E}=\frac{\kappa}{2\pi}\delta A$, where $\kappa$ is the surface gravity and $A$ is the surface area of the horizon\footnote{It is instructive to refer to Ref.~\cite{Poisson} for a more enlightening step-by-step derivation of the formula which shows explicitly how surface gravity factors out in the formula.}. This identity becomes identical to the first law of thermodynamics, $\delta\mathcal{E}=T\delta S$, provided only that one identifies --- up to proportionality factors --- temperature with the surface gravity $\kappa$ and entropy with the surface area $A$ of the horizon. On the other hand, provided that $\pounds_\xi\phi=0$ for all the fields of the theory, one has for a given hypersurface $\Xi$ the identity, $\int_{\partial\Xi}\delta{\bf Q}[\xi]-i_\xi\boldsymbol{\Theta}=0$ \cite{IyerWald}. This, and the fact that the Killing vector $\xi^\mu$ (which coincides with the timelike Killing field $t^\mu$ at spatial infinity) vanishes on the bifurcation 2-surface $\Sigma$ of the horizon, leads in GR to $\delta\mathcal{E}=\delta\int_{\Sigma}{\bf Q}[\xi]$ \cite{IyerWald}, provided that the only other (interior) boundary of the spatial hypersurface $\Xi$ of interest is the apparent horizon of the black hole. Comparing now this identity with the first law, one deduces that the black hole entropy should be given by $S=\frac{2\pi}{\kappa}\int_\Sigma{\bf Q}[\xi]$, provided that one is able to show that the variation $\delta\int_{\Sigma}{\bf Q}[\xi]$ can indeed be expressed in the form of a product of $\frac{\kappa}{2\pi}$ times a term of the form $\delta S$, for some scalar $S$. It turns out that this is indeed the case in GR, whence the Wald prescription, $S=\frac{2\pi}{\kappa}\int_\Sigma{\bf Q}[\xi]$ \cite{WaldEntropy,IyerWald}.

Now, since the spacetime dynamics in GR is equivalent to the dynamics described by TEGR, the first law of black hole mechanics is obviously preserved in the framework of TEGR. On the other hand, combining our results (\ref{tTheta}) and (\ref{ADM}) with the fact that the Killing vector $\xi^\mu$ coincides with the timelike Killing field $t^\mu$ at spatial infinity, as well as the fact that, $\int_{\partial\Xi}\delta{\bf Q}[\xi]-i_\xi\boldsymbol{\Theta}=0$, guaranteed by, $\pounds_\xi\phi\equiv\pounds_\xi e^a_\mu=0$, we deduce, as in GR, that $\delta\mathcal{E}=\delta\int_{\Sigma}{\bf Q}[\xi]$. Unlike in GR, however, the 2-surface $\Sigma$ of the horizon does not have to be a bifurcation 2-surface on which the Killing vector vanishes.

We now use the black hole's first law to infer that entropy should be extracted using the inverse of $\frac{\kappa}{2\pi}$ by writing, $\delta S=\frac{2\pi}{\kappa}\delta\int_\Sigma{\bf Q}[\xi]=\delta\int_\Sigma\frac{2\pi}{\kappa}{\bf Q}[\xi]$. In the second step we moved the surface gravity $\kappa$ inside the integral based on the constancy of $\kappa$ over the entire horizon as well as its insensitivity to the variation $\delta\phi$ of the dynamical fields of the theory.
It is worth noting here that the issue of moving the surface gravity around in the first order formalism of GR has already been pointed out in Ref.~\cite{Corichi}. 

In Ref.~\cite{Corichi}, $\kappa$ had to be moved outside the integral to make the first law hold, precluding thus {\it a priori} Wald's approach from being sufficiently general as to relate black hole entropy to Noether charge in the first order formalism of GR. The argument given in Ref.~\cite{Corichi} was that $\kappa$ depends on the surface area of the horizon \cite{Ashtekar}. We did move here $\kappa$ inside the integral for, as pointed out also in Ref.~\cite{Ashtekar}, although the surface gravity does depend on the surface area, it does not depend on the ``shape" of the latter. In the derivation of the first law of black hole mechanics in the second order formalism of GR, one indeed assumes an adiabatic process that does not affect $\kappa$ (see Ref.~\cite{Poisson}). This can be thought of as a series of infinitesimal changes in the shape of the horizon. 

However, since our argument does not require putting $\kappa$ inside the integral, for the sake of generality, we are going to keep the former outside the latter. It follows then that the black hole entropy in TEGR can be expressed as,
\begin{align}\label{TEGREntropy}
    S&=\frac{2\pi}{\kappa}\int_\Sigma{\bf Q}[\xi]\nonumber\\
    &=\frac{1}{8\kappa}\int_\Sigma\xi^a{\mathcal{S}_a}^{\alpha\beta}\boldsymbol{\epsilon}_{\alpha\beta\gamma\lambda}\nonumber\\
    &=\frac{1}{4\kappa}\int_V{\rm d}^3x\,\partial_i\left(e\xi^a{\mathcal{S}_a}^{0i}\right).
\end{align}
In the last step we have used again Stokes' theorem and integrated over the volume $V$ bounded by the black hole horizon. This result shows that entropy, like the ADM mass, can very naturally be expressed as a volume integral. Therefore, the integrand in the last line can be interpreted as an entropy density. 

It is worth noting here that in Refs.~\cite{TorsionEntropy1,QuasiAndTorsion,WeiNing} it is found that torsion does not contribute to black hole entropy in Riemann-Cartan spacetimes (see, however, Ref.~\cite{Miao}). In such spacetimes torsion plays the role of an independent degree of freedom conditioned by the presence of matter with nonzero intrinsic spin. We do get in TEGR a contribution to entropy from torsion because, unlike in other theories of gravity with torsion, the latter becomes in TEGR a substitute for curvature. 

We would like to stress here again the important fact that nowhere did we have to invoke the bifurcation 2-surface of the horizon and the vanishing of the Killing vector on the latter. Formula (\ref{TEGREntropy}) works in fact on any cross section of the apparent horizon. Thus, in agreement with what was found in Ref.~\cite{Corichi} within the first order formalism of GR, extracting entropy in TEGR does not necessitate a bifurcate horizon either. This fact actually constitutes another advantage of adopting Wald's approach in TEGR, for it is well known that more realistic black holes emerge from gravitational collapse, the spacetime of which is not expected to possess any bifurcation surface. 

This departure from the restrictions of the second-order formalism of GR is made possible by the use of the tetrad fields in the first-order formalism. In fact, besides making TEGR a first-order formalism, the tetrad formalism yields a symplectic potential that depends on the variation of the tetrad fields themselves, as opposed to what happens within GR where the variation of the first derivatives of the metric is required. As such, the mere invariance of the tetrad fields replaces the requirement to have a vanishing Killing field. Physically, this could be viewed as if, unlike the first derivatives of the metric in GR, the tetrad fields in TEGR are sensitive even to a smooth horizon without a bifurcation surface.

Now, to evaluate the surface integral in Eq.~(\ref{TEGREntropy}), we proceed as follows. First, we use the identity $\epsilon^{\mu\nu}=\epsilon^{\mu\nu\rho\sigma}N_\rho\xi_\sigma$, where $N^\mu$ is an auxiliary null vector on the horizon, normalized such that $N_\mu\xi^\mu=-1$ and $\boldsymbol{\epsilon}_{\mu\nu}$ is the volume element on the 2-surface $\Sigma$ \cite{Wald}. With this identity, the surface integral in Eq.~(\ref{TEGREntropy}) yields,
\begin{align}\label{EntropyStep1}
    S&=\frac{1}{16\kappa}\int_\Sigma\xi^a{\mathcal{S}_a}^{\alpha\beta}\left(\boldsymbol{\epsilon}^{\mu\nu}\boldsymbol{\epsilon}_{\mu\nu\alpha\beta}\right)\boldsymbol{\epsilon}_{\gamma\lambda}\nonumber\\
    &=\frac{1}{16\kappa}\int_\Sigma\xi^a{\mathcal{S}_a}^{\alpha\beta}\left(\boldsymbol{\epsilon}^{\mu\nu\rho\sigma}N_\rho\xi_\sigma\boldsymbol{\epsilon}_{\mu\nu\alpha\beta}\right)\boldsymbol{\epsilon}_{\gamma\lambda}\nonumber\\
    &=\frac{1}{4\kappa}\int_\Sigma\xi^a{\mathcal{S}_a}^{\alpha\beta}N_\beta\xi_\alpha\boldsymbol{\epsilon}_{\gamma\lambda}.
\end{align}
In the last step we have used $\boldsymbol{\epsilon}^{\mu\nu\rho\sigma}\boldsymbol{\epsilon}_{\mu\nu\alpha\beta}=-4\delta^{[\rho}_\alpha\delta^{\sigma]}_\beta$ \cite{Wald}.

Next, a straightforward calculation shows that $\pounds_\xi g_{\mu\nu}=\nabla_\mu\xi_\nu+\nabla_\nu\xi_\mu-\xi^\rho\left(T_{\mu\nu\rho}+T_{\nu\mu\rho}\right)$. This means that for a Killing vector $\xi^\mu$, for which one has $\pounds_\xi g_{\mu\nu}=0$, the usual Killing equation ($\nabla_\mu\xi_\nu+\nabla_\nu\xi_\mu=0$) should be replaced by the modified Killing equation, $\nabla_\mu\xi_\nu+\nabla_\nu\xi_\mu=\xi^\rho\left(T_{\mu\nu\rho}+T_{\nu\mu\rho}\right)$ \cite{DeyLiberatiPranzetti}. Contracting both sides of this equation with $\xi^\mu$, and using the antisymmetry of the torsion tensor in its last two indices gives,
\begin{equation}\label{TEGRSurfaceGravity}
    \xi^\mu\xi^\rho T_{\mu\rho\nu}=\xi^\mu\nabla_\mu\xi_\nu+\xi^\mu\nabla_\nu\xi_\mu.
\end{equation}
Maintaining the requirement that surface gravity $\kappa$ be unique and be defined as usual by $\xi^\mu\nabla_{\mu}\xi_\nu=\kappa\xi_\nu$ as well as $\xi^\mu\nabla_{\nu}\xi_\mu=-\kappa\xi_\nu$, the authors in Ref.~\cite{DeyLiberatiPranzetti} argued that in spacetimes with torsion the identity $\xi^\mu\xi^\nu T_{\mu\nu\rho}=0$ should hold. Keeping the same requirement here in TEGR, let us expand $\xi^\rho T_{\mu\rho\nu}$ into $\xi^\rho T_{\mu\rho\nu}=pg_{\mu\nu}+q\xi_\mu N_\nu+s\xi_\nu N_\mu$, for some scalars $p$, $q$ and $s$ to be determined. Contracting both sides of this identity separately with $\xi^\mu$ and $\xi^\nu$, and using the antisymmetry of torsion as well as $N_\mu\xi^\mu=-1$, we deduce that $p=q=s$. Therefore, by contracting this time both sides of the identity with $g^{\mu\nu}$, we conclude that $\xi^\mu T_\mu=2p$.

Finally, using the definition of the superpotential ${\mathcal{S}_a}^{\mu\nu}$ in terms of torsion as given below Eq.~(\ref{Torsion+Contortion}), together with the requirements $\xi^\mu\xi^\nu T_{\mu\nu\rho}=0$ and $\xi^\mu T_\mu=2p$ we just deduced, we find,
\begin{align}\label{EntropyStep2}
    S&=\frac{1}{4\kappa}\int_\Sigma\xi^a{\mathcal{S}_a}^{\alpha\beta}N_\beta\xi_\alpha\boldsymbol{\epsilon}_{\gamma\lambda}\nonumber\\
    &=\frac{1}{4\kappa}\int_\Sigma\xi^\rho({T_\rho}^{\alpha\beta}-\tfrac{1}{2}{T^{\alpha\beta}}_\rho-\delta^\beta_\rho T^\alpha+\delta^\alpha_\rho T^\beta)N_\beta\xi_\alpha\boldsymbol{\epsilon}_{\gamma\lambda}\nonumber\\
    &=\frac{2pA}{4\kappa}.
\end{align}
We see that the value $p=\kappa/2$ yields the familiar area law for entropy. To check that $\xi^\mu T_\mu=\kappa$ is indeed what one recovers in TEGR, let us examine a concrete example. Let us use the metric of the Schwarzschild black hole of mass $m$, given by $g_{00}=1-2m/r=-1/g_{11}$. One easily extracts the tetrad fields for this metric, from which one finds the nonvanishing components of torsion to be, $T^0\,\!_{10}={g_{00}}_{,r}/2g_{00}$ and $T^2\,\!_{12}=T^3\,\!_{13}=(1-\sqrt{-g_{11}})/r$  \cite{SchwarzschildTorsion}. Computing then the only nonvanishing trace component $T_1$ and using the outgoing null normal $(1,g_{00},0,0)$ that gives the right orientation for the 2-surface $\Sigma$ on which it coincides with the null Killing vector, we easily evaluate the contraction $\xi^\mu T_\mu$ on the horizon and find $m/r^2=1/4m$, which is just the surface gravity of the Schwarzschild black hole.
\section{Behavior under Weyl Transformation}\label{sec:UnderWeyl}
The issue that arises when applying conformal transformations in GR to the Wald entropy is the fact that the latter is found to be invariant under conformal transformations whereas the familiar area law suggests that entropy would transform like a surface area \cite{FebruaryPaper}. Now that we saw that entropy in TEGR is more naturally expressed as a volume integral, the question that arises is whether such an expression would still make entropy invariant under Weyl transformations as in GR. If so, a second question would then necessarily arise. One would then indeed want to know whether such an invariance is compatible with the interpretation of the integrand in the result (\ref{TEGREntropy}) as an entropy density.

A Weyl conformal transformation consists in rescaling the metric with an everywhere regular and positive factor $\Omega^2(x)$. Formally, this reads, $\tilde{g}_{\mu\nu}=\Omega^2(x)g_{\mu\nu}$. The new spacetime obtained by such a transformation is usually called the conformal frame, or the Einstein frame, as opposed to the original spacetime called usually the Jordan frame \cite{ConformalReference}. In order to extract the black hole entropy in the conformal frame, we need first to find the expression of the new gravitational Lagrangian and then apply to it Wald's algorithm to extract the Noether charge.

Under the Weyl rescaling of the metric, the tetrads transform as $\tilde{e}^a_\mu=\Omega e^a_\mu$. Thereby, the transformation of the remaining terms of the TEGR Lagrangian are easily found to be, $\tilde{e}=\Omega^{4}e$, $\tilde{e}^\mu_a=\Omega^{-1}e^{\mu}_a$, $\tilde{T}^\rho\,\!_{\mu\nu}=T^\rho\,\!_{\mu\nu}+\delta^\rho_\mu\partial_\nu\Omega-\delta^\rho_\nu\partial_\mu\Omega$ and $\tilde{\mathcal{S}}_\rho\,\!^{\mu\nu}=\Omega^{-2}\mathcal{S}_{\rho}\,\!^{\mu\nu}$. Notice that to obtain the transformation of torsion, we assumed, as is usually done in TEGR \cite{ConfTEGR1,ConfTEGR2,ConfTEGR3}, that the spin connection is conformally invariant. This is due to the fact that in TEGR the spin connection is taken to be a purely inertial Lorentz connection, totally unaffected by the rescaling of the spacetime metric. The transformation of torsion in TEGR is thus similar to what is usually found in the literature on the search for conformal invariance in the more general Riemann-Cartan spacetimes \cite{ConformalTorsion1,ConformalTorsion2,ConformalTorsion3,ConformalTorsion4}. In those spacetimes, however, the spin connection does not have to be restricted to the inertial Lorentz connection and, hence, it is {\it a priori} expected to be affected by the Weyl transformation as well. 
Using now these transformations, the TEGR Lagrangian within the action (\ref{TPAction}) takes the following form in the conformal frame:
\begin{equation}\label{ConfLagrangian}
    \tilde{{\bf{L}}}=\frac{\tilde{\boldsymbol{\epsilon}}}{32\pi}\left(\frac{\tilde{\mathbb{T}}}{\Omega^2}+\frac{8\tilde{T}^{\mu}}{\Omega^3}\tilde{\nabla}_\mu\Omega-\frac{12}{\Omega^4}\tilde{\nabla}_\mu\Omega\tilde{\nabla}^\mu\Omega\right).
\end{equation}

Before we proceed to the extraction of the symplectic potential, it is important to pause here and notice that the first and the last terms of this Lagrangian are very reminiscent of Brans-Dicke's scalar-tensor generalization of GR. By the redefinition $\Omega^{-2}=\phi$, the scalar field $\phi$ would indeed play a role analogous to the Brans-Dicke scalar field. In GR, one in fact goes from a conformally transformed Einstein-Hilbert Lagrangian to a real Brans-Dicke Lagrangian by replacing the anomalous Brans-Dicke parameter $-3/2$ that results from such a procedure by an arbitrary parameter $\omega(\phi)$ (see the discussion in Ref.~\cite{FDV}). This technique works actually even for highly nonlinear models such as the one proposed  in Refs.~\cite{VPL1,VPL2,VPL3}. The structure of the Lagrangian (\ref{ConfLagrangian}) represents thus a potential prototype for general scalar-tensor teleparallel gravity theories, provided only that both constant factors $8$ and $-12$ in it be replaced by arbitrary functions of the field $\phi$. We therefore conclude that, unlike the Lagrangians introduced in Refs.~\cite{TPDark,TPQuintessence,BDTP,TPPhi} to be used in scalar-tensor (or even ``Brans-Dicke") teleparallel gravity theories, these extensions of teleparallel gravity would be more accurately described by letting their Lagrangian acquire a coupling between the gradient of the scalar field and torsion.

Let us now proceed with the extraction of the symplectic potential. As in the original frame, the variation of this Lagrangian takes the simple compact form, $\delta\tilde{{\bf L}}={\tilde{{\bf E}}_a}\,\!^\mu\,\delta \tilde{e}^a_\mu+\tilde{\bf E}^{(\Omega)}\,\delta\Omega+{\rm d}\tilde{{\boldsymbol\Theta}}$. However, the new tetrad and $\Omega$ equations of motion, given, respectively, by ${\tilde{{\bf E}}_a}\,\!^\mu=0$ and $\tilde{\bf E}^{(\Omega)}=0$, as well as the new symplectic potential $\tilde{{\boldsymbol\Theta}}$ are all much more involved than those that arise in the original frame. In fact, the explicit expression of the tetrad equations of motion are given by Eq.~(\ref{ConfE}). Using these, the symplectic potential is easily extracted and is found to be given by Eq.~(\ref{ConfSymplectic}). The explicit expression of the Noether charge we then obtain from such a symplectic potential is as given in Eq.~(\ref{ConfCharge}). Using the latter, the entropy of the black hole in the conformal frame can be computed\footnote{Note that the requirement $\pounds_{\tilde\xi}\tilde{e}^a_\mu=0$ that guarantees the equation $\int_{\partial\tilde{\Xi}}\delta{\tilde{\bf Q}}[\tilde{\xi}]-i_{\tilde{\xi}}\tilde{\boldsymbol{\Theta}}=0$ to hold in the conformal frame is now augmented by the requirement to also have $\pounds_{\tilde{\xi}}\Omega=\tilde{\xi}^\mu\tilde{\nabla}_\mu\Omega=0$. The latter is actually already guaranteed by the fact that $\tilde{\xi}^\mu\tilde{\nabla}_\mu\Omega=0$ is also the requirement for a Killing vector field to exist in the conformal frame \cite{AugustPaper}.} and expressed in terms of the entropy of the original frame as follows:
\begin{align}\label{ConfTEGREntropy}
    \tilde{S}&=\frac{2\pi}{\tilde{\kappa}}\int_{\tilde{\Sigma}}{\bf \tilde{Q}}[\tilde{\xi}]\nonumber\\
    &=\frac{1}{8\tilde{\kappa}}\int_{\tilde{\Sigma}}\frac{\tilde{\xi}^a}{\Omega^2}\left({\tilde{\mathcal{S}}_a}\,\!^{\alpha\beta}-2\Omega^2\tilde{\Sigma}_a\,\!^{\alpha\beta}\right)\tilde{\boldsymbol{\epsilon}}_{\alpha\beta\gamma\lambda}\nonumber\\
    &=\frac{1}{8\tilde{\kappa}}\int_{\tilde{\Sigma}}\frac{\tilde{\xi}^a}{\Omega^2}{\tilde{\mathcal{S}}_a}^{\alpha\beta}\tilde{\boldsymbol{\epsilon}}_{\alpha\beta\gamma\lambda}\nonumber\\
    &=\frac{1}{8\kappa}\int_{\tilde{\Sigma}}\xi^a{\mathcal{S}_a}^{\alpha\beta}\boldsymbol{\epsilon}_{\alpha\beta\gamma\lambda}\nonumber\\
    &=S.
\end{align}
To obtain the third line, we have used the fact that, as shown in Ref.~\cite{AugustPaper}, the existence of the Killing vector field $\tilde{\xi}^\mu$ in the conformal frame is conditioned by having the conformal factor $\Omega$ satisfy also $\tilde{\xi}^{[\mu}\tilde\nabla^{\nu]}\Omega=0$, and hence, like surface gravity, $\Omega$ is uniform all over the horizon. The contribution of the last term in the second line is indeed proportional to $\tilde{\xi}^{[\mu}\tilde{\nabla}^{\nu]}\Omega$, as can be seen by computing the contraction $\tilde{\xi}^a\tilde{\Sigma}_a\,\!^{\mu\nu}$ using our definition of the induced torsion $\tilde{\Sigma}_a\,\!^{\mu\nu}$ given below Eq.~(\ref{ConfE}). In the fourth line we have used the fact that under a Weyl transformation of the metric, the Killing vector field transforms into $\tilde{\xi}^\mu=\xi^\mu/\Omega$ and the surface gravity transforms into $\tilde{\kappa}=\tilde{\kappa}/\Omega$ \cite{AugustPaper}.

We clearly see then that, as was the case in GR, entropy is invariant under Weyl's conformal transformations. Being reducible here to a volume integral, however, it is easier to understand the origin of such an invariance. Indeed, in this case it is simply due to the fact that one now is also able to integrate over a volume an entropy density which does transform like $({\rm{volume}})^{-1}$.
\section{Summary \& Discussion}\label{sec:Conclusion}
Black hole thermodynamics has been investigated within the framework of TEGR based on Wald's algorithm for diffeomorphism invariant theories of gravity. Our result shows that, unlike what is found within the framework of GR, black hole entropy is more naturally expressed as a volume integral. This result makes a perfect parallel with the already well-known fact that within TEGR the ADM mass associated to the gravitational energy is also expressed as a volume integral. This could actually be understood in a natural way, both mathematically and physically, as follows.

Recall, indeed, that geometrically TEGR manages to express the ADM mass as a volume integral thanks to the use of the tetrad field, which is geometrically richer than the metric field in the sense that the former represents the square root of the latter. The consequence of this, as we saw in the Introduction and in Appendix ~\ref{A}, is that TEGR is a first-order theory, both in its Lagrangian and in its equations of motion. Thus, what was second order within GR became first order within TEGR. In other words, thanks to the tetrad field, one is, in some sense, able to ``integrate" out GR to get TEGR. This simple pattern is recovered for the case of the ADM mass. Indeed, the latter within GR is a surface integral that is equivalent to a quantity that appears within TEGR as an integral over a volume of a total derivative. According to this logic then, we naturally expect that entropy within GR should be equivalent to a quantity that consists of an integral of a volume within TEGR, i.e., an integral of an entropy density. In other words, what within GR appeared as a mere surface term, TEGR has been able to resolve it to reveal it to be actually spread over a volume. The miracle thus rests mathematically on the use of a richer structure provided by the frame fields.

On the other hand, such a miracle can in fact be understood physically as well. As alluded to in the Introduction, this parallel within the framework of TEGR between the fate of the gravitational energy and that of black hole entropy actually has a common origin. Indeed, as already recalled there, even the very existence of an energy-momentum density tensor for the gravitational field is due to the possibility of filtering out inertial effects from pure gravity --- the contribution of the former being of a pseudotensorial nature, in contrast to the contribution of the latter. In this sense, it is not surprising that black hole entropy becomes also subjected within the framework of TEGR to the same fate. Recall indeed that, thanks to the strong equivalence principle, the similarity between the black hole entropy area law and the area law of the entanglement entropy that arises in the reference frame of an accelerated observer becomes less obscure. This very observation can, however, be turned upside down by arguing that, after eliminating inertial effects from our description of the dynamics of spacetime, the behavior of black hole entropy does not necessarily have to conform to what is measured by an accelerated observer.

The motivation behind the investigation conducted in this paper came actually from the other curious fact that, within the framework of GR, Wald's approach yields a conformally invariant entropy, in contrast to what one expects from an area law. It was argued in Ref.~\cite{FebruaryPaper} that the fundamental reason for such a behavior is due to the fact that Wald's approach is not so much about a geometric {\it relation} within the theory than about a geometric {\it property} of the theory. Here, the geometric property is diffeomorphism invariance and the latter is preserved under Weyl's transformations. In fact, no mixing between matter and geometry is then ever required as is the case with some of the geometric concepts of mass in GR \cite{Hammad2,Hammad3}, with Einstein equations \cite{ConformalReference} and with the recipe \cite{JacobsonPRL} for extracting Einstein equations from Clausius' thermodynamic relation \cite{FebruaryPaper}. Therefore, regardless of the separability between inertia and gravity within the framework of TEGR, the latter is still diffeomorphism invariant and, hence, is also expected to provide an entropy that is invariant under Weyl's transformation of spacetime. Our result in Sec.~\ref{sec:Entropy} confirms just this expectation. In addition to this, however, one gains now a nice new interpretation of this invariance that was not possible within GR. The invariance can in fact simply be interpreted as being due to the fact that one is integrating an entropy density over a volume.

Finally, we would like to conclude this section with a brief interesting speculation. It was argued in Ref.~\cite{TEGR} that the quantization problem of gravity within GR could be cured if one uses TEGR instead because precisely of this crucial separation between gravity and inertia as well as its gauge formulation (see Ref.~\cite{TEGRGauge} for a more recent technical discussion on this last point). See also the works \cite{TEGRQG1,TEGRQG2} and the references therein for recent progress on quantization and on quantum cosmology in TEGR. It is then not excluded that this work could also lead to a better understanding at the quantum level of black hole entropy and even help resolve some of the deepest issues posed by the latter. 

\section*{Acknowledgments}
F.H., D.D and A.T.-R are supported by the Natural Sciences and Engineering Research
Council of Canada (NSERC) Discovery Grant (RGPIN-2017-05388). D.B. is supported by the Fonds de Recherche du Qu\'ebec - Nature et Technologies (FRQNT) Grant (280517).

\appendix
\section{Field equations of TEGR}\label{A}
In this Appendix we display the details of the derivation of the field equations in TEGR. This derivation will indeed play an important role in the extraction of the symplectic potential $\boldsymbol{\Theta}$ in Appendix \ref{B}. First, varying the Lagrangian yields,
\begin{align}\label{AppVaryL}
\delta{\bf L}&=\frac{1}{32\pi}\delta\left( \boldsymbol{\epsilon}\mathbb{T}\right)\nonumber\\
&=\frac{\boldsymbol{\epsilon}}{32\pi e}\left[\frac{\partial(e \mathbb{T})}{\partial e^a_\mu}\delta e^a_\mu+\frac{\partial(e\mathbb{T})}{\partial(\partial_\nu e^a_\mu)}\delta(\partial_\nu e^a_\mu)\right]\nonumber\\
&=\frac{\boldsymbol{\epsilon}}{32\pi e}\left[\frac{\partial(e \mathbb{T})}{\partial e^a_\mu}-\partial_\nu\frac{\partial(e\mathbb{T})}{\partial(\partial_\nu e^a_\mu)}\right]\delta e^a_\mu\nonumber\\
&\quad+\frac{\boldsymbol{\epsilon}}{32\pi e}\partial_\nu\left[\frac{\partial(e\mathbb{T})}{\partial(\partial_\nu e^a_\mu)}\delta e^a_\mu\right].
\end{align}
When searching for the field equations, one discards the total derivative of the last line as being a boundary term that does not contribute in the variation. Therefore, the field equations of TEGR take on the elegant and simple form, $\partial_{\nu}(e\mathcal{S}_a\,\!^{\mu\nu})=8\pi e\mathcal{J}_a^{\mu}$, with
\begin{align}\label{AppEOM}
-32\pi e\mathcal{J}_a^\mu&\equiv\frac{\partial(e\mathbb{T})}{\partial e^a_\mu}=ee^\mu_a\mathbb{T}+e\frac{\partial\mathbb{T}}{\partial e^a_\mu}\nonumber\\
&=e\left(e^\mu_a\mathbb{T}+4e^\rho_a\,{\mathcal{S}_b}^{\mu\nu}{T^b}_{\nu\rho}-4{\omega_\nu}^b\,\!_{a}{\mathcal{S}_b}^{\mu\nu}\right),\nonumber\\
-4e\,{\mathcal{S}_a}^{\mu\nu}&=\frac{\partial(e\mathbb{T})}{\partial(\partial_\nu e^a_\mu)}.
\end{align}
The tensor ${\mathcal{S}_a}^{\mu\nu}$ is called the ``superpotential'' and ${\mathcal{J}_a}^\mu$ is called the ``gravitational energy-momentum density'' or simply the ``gravitational current density'' \cite{TEGR,MalufReview}.
\section{The symplectic potential and Noether charge in TEGR}\label{B}
With the above equations of motion at hand we can now extract the symplectic potential by following a simple algorithm \cite{WaldEntropy,IyerWald}. First, by varying the Lagrangian, we saw from Eq.~(\ref{AppVaryL}) that besides the equations of motion, there is the extra term of the last line which is a four divergence. Therefore, using the definitions (\ref{AppEOM}) for the various derivatives, we have the following expression for the variation of the Lagrangian, 
\begin{align}\label{ThetaTP}
\delta{\bf L}&=\frac{\,\boldsymbol{\epsilon}}{e}\!\left[-e{\mathcal{J}_a}^\mu+\frac{1}{8\pi}\partial_\nu(e{\mathcal{S}_a}^{\mu\nu})\right]\!\delta e^a_\mu\!-\!\frac{\,\boldsymbol{\epsilon}}{8\pi e}\partial_\nu\left(e{\mathcal{S}_a}^{\mu\nu}\delta e_\mu^a\right)\nonumber\\
&={{\bf E}_a}^\mu\,\delta e^a_\mu+{\rm d}{\boldsymbol\Theta}.
\end{align}
Here, we have introduced the 4-form ${{\bf E}_a}^\mu$, the vanishing of which yields the equations of motion as can be seen from the content of the square brackets of the first line. On the other hand, the four-divergence structure of the last term allows us to turn the latter into an exact form ${\rm d}\boldsymbol\Theta$, where the symplectic potential 3-form $\boldsymbol{\Theta}$ can directly be read off, $\boldsymbol{\Theta}_{\beta\gamma\lambda}=-(8\pi)^{-1}{\mathcal{S}_a}^{\mu\alpha}\delta e_\mu^a\boldsymbol{\epsilon}_{\alpha\beta\gamma\lambda}$.

Next, using this general symplectic 3-form $\boldsymbol{\Theta}$, one builds the current 3-form ${\bf J}=\boldsymbol{\Theta}-i_\xi{\bf L}$ in which the arbitrary variation $\delta e^a_\mu$ of the tetrad field is replaced by its Lie derivative $\pounds_\xi e^a_\mu$ along an arbitrary vector field $\xi^{\mu}$. Such a current 3-form $\bf J$ becomes then closed on shell, {\it i.e.}, when the equations of motion are satisfied: ${\bf E}_a\,\!^\mu=0$. In fact, we have in this case, 
\begin{align}\label{dJ}
    {\rm d} {\bf J}&={\rm d}\boldsymbol{\Theta}-{\rm d}(i_\xi{\bf L})\nonumber\\
    &={\rm d}\boldsymbol{\Theta}-\pounds_\xi{\bf L}+i_\xi{\rm d}{\bf L}\nonumber\\
    &={\rm d}\boldsymbol{\Theta}-\pounds_\xi{\bf L}\nonumber\\
    &={\rm d}\boldsymbol{\Theta}-{{\bf E}_a}^\mu\pounds_\xi e^ a_\mu-{\rm d}\boldsymbol{\Theta}\nonumber\\
    &=-{{\bf E}_a}^\mu\pounds_\xi e^ a_\mu\nonumber\\
    &=0.\quad (\text{on shell})
\end{align}
In the second line, use has been made of Cartan's formula for differential forms, $\pounds_\xi{\bf L}=i_\xi{\rm d}{\bf L}+{\rm d}(i_\xi{\bf L})$. The last line implies that there must exist, on shell, a 2-form $\bf Q$ such that ${\bf J}={\rm d}\bf{Q}$. 

Our task then is to find out what this 2-form $\bf Q$ is. The strategy then consists in building the 3-form $\bf J$ and checking if it can really be written as an exterior derivative of a 2-form. First, after replacing in Eq.~(\ref{Symplectic}) the variation $\delta e^a_\mu$ by the generalized Lie derivative $\mathcal{L}_\xi e^a_\mu=\pounds_\xi e^a_\mu+\mathcal{M}^a\,\!_b e^b_\mu$, as described in the text below Eq.~(\ref{Symplectic}), the latter acquires the following explicit form in terms of the Killing vector field $\xi^\mu$:
\begin{align}\label{Theta1stStep}
    \boldsymbol{\Theta}_{\beta\gamma\lambda}&=-\frac{1}{8\pi}{\mathcal{S}_a}^{\mu\alpha}\left(\xi^\rho\partial_\rho e^a_\mu+e^a_\rho\partial_\mu\xi^\rho+{\mathcal{M}^a}_b e^b_\mu\right)\boldsymbol{\epsilon}_{\alpha\beta\gamma\lambda}\nonumber\\
    &=-\frac{1}{8\pi}{\mathcal{S}_a}^{\mu\alpha}\left(\xi^\rho{T^a}_{\rho\mu}+\partial_\mu\xi^a+{\omega_\mu}^a\,_b \xi^b\right)\boldsymbol{\epsilon}_{\alpha\beta\gamma\lambda}\nonumber\\
    &\quad+\frac{1}{8\pi}{\mathcal{S}_a}^{\mu\alpha}e^b_\mu\left(\xi^\rho{\omega_\rho}^a\,_b-{\mathcal{M}^a}_b\right)\boldsymbol{\epsilon}_{\alpha\beta\gamma\lambda}\nonumber\\
    &=-\frac{1}{8\pi}{\mathcal{S}_a}^{\mu\alpha}\,{T^a}_{\rho\mu}\xi^\rho\,\boldsymbol{\epsilon}_{\alpha\beta\gamma\lambda}-\frac{1}{8\pi}\partial_\mu\left({\mathcal{S}_a}^{\mu\alpha}\,\xi^a\boldsymbol{\epsilon}_{\alpha\beta\gamma\lambda}\right)\nonumber\\
    &\quad+\frac{1}{8\pi }\left[e^{-1}\partial_\mu\left(e{\mathcal{S}_a}^{\mu\alpha}\right)\xi^a-{\mathcal{S}_a}^{\mu\alpha}{\omega_\mu}^a\,_b \xi^b\right]\boldsymbol{\epsilon}_{\alpha\beta\gamma\lambda}\nonumber\\
    &\quad+\frac{1}{8\pi}{\mathcal{S}_a}^{\mu\alpha}\left(\xi^\rho{\omega_\rho}^a\,_b-{\mathcal{M}^a}_b\right)e^b_\mu\boldsymbol{\epsilon}_{\alpha\beta\gamma\lambda}.
\end{align}
In the first line we have introduced the generalized Lie derivative, in the second equality we introduced the torsion tensor (\ref{Torsion+Contortion}), and in the last equality we have integrated by parts. Next, thanks to the equations of motion, the third term in the third equality can be traded for the gravitational current $e\mathcal{J}_a^\mu$. The latter can, in turn, be replaced by its full explicit expression as given in the first line of Eq.~(\ref{AppEOM}). The expression (\ref{Theta1stStep}) then simplifies greatly and reduces to, 
\begin{align}
    \boldsymbol{\Theta}_{\beta\gamma\lambda}&=-\frac{1}{8\pi}\partial_\mu\left(\xi^a{\mathcal{S}_a}^{\mu\alpha}\boldsymbol{\epsilon}_{\alpha\beta\gamma\lambda}\right)+\frac{1}{32\pi}\mathbb{T}\xi^\alpha\boldsymbol{\epsilon}_{\alpha\beta\gamma\lambda}\nonumber\\
    &\quad+\frac{1}{8\pi}{\mathcal{S}_a}^{\mu\alpha}\left[\xi^\rho{\omega_\rho}^a\,_b-{\mathcal{M}^a}_b \right]e^b_\mu\,\boldsymbol{\epsilon}_{\alpha\beta\gamma\lambda}.
\end{align}
Using this last expression of $\boldsymbol{\Theta}$, we can now build the current 3-form $\bf J$ as follows:
\begin{align}\label{Current}
{\bf J}_{\beta\gamma\lambda}&=\boldsymbol{\Theta}_{\beta\gamma\lambda}-(i_\xi{\bf L})_{\beta\gamma\lambda}\nonumber\\
    &=-\frac{1}{8\pi}\partial_\mu\left(\xi^a{\mathcal{S}_a}^{\mu\alpha}\boldsymbol{\epsilon}_{\alpha\beta\gamma\lambda}\right)\nonumber\\
    &\quad+\frac{1}{8\pi}{\mathcal{S}_a}^{\mu\alpha}\left[\xi^\rho{\omega_\rho}^a\!\,_b-{\mathcal{M}^a}_b\right]e^b_\mu\boldsymbol{\epsilon}_{\alpha\beta\gamma\lambda}.
\end{align}
We clearly see then that, in order for the 3-form current ${\bf J}_{\beta\gamma\lambda}$ to be an exact differential 2-form, as required by the existence of the Noether charge $\bf Q$, the Lorentz rotation matrix ${\mathcal{M}^a}_b(x,\xi)$ should be given by:
\begin{equation}
    {\mathcal{M}^a}_b(x,\xi)=\xi^\rho{\omega_\rho}^a\,_b=\xi^\rho(\Lambda^{-1})^a\,_c\,\partial_\rho\Lambda^c\,_b.
\end{equation}
In this case, the current 3-form $\bf J$ indeed becomes an exact form ${\rm d}{\bf Q}$, where the charge 2-form $\bf Q$ reads,
\begin{equation}
    {\bf Q}_{\gamma\lambda}=\frac{1}{16\pi}\xi^a{\mathcal{S}_a}^{\alpha\beta}\boldsymbol{\epsilon}_{\alpha\beta\gamma\lambda}.
\end{equation}
Notice that, as explained in the text, by working in the class of reference frames in which the spin connection vanishes globally, one does not need to introduce the generalized Lie derivative $\mathcal{L}_\xi$. The usual Lie derivative $\pounds_\xi e^ a_\mu$ would then be amply sufficient.
\section{Field equations, symplectic potential and Noether charge in the conformal frame}\label{C}
Under the Weyl transformation $\tilde{g}^{\mu\nu}=\Omega^2(x)g_{\mu\nu}$, with a spacetime-dependent conformal factor $\Omega(x)$, the TEGR Lagrangian takes the form (\ref{ConfLagrangian}). The arbitrary variation of the fields in the latter then yields,
\begin{equation}\label{ConfVariedLagrangian}
    \delta\tilde{\bf L}=\tilde{\bf E}_a\,\!^\mu\delta\tilde{e}^a_\mu+\tilde{\bf E}^{(\Omega)}\delta\Omega+{\rm d}\tilde{\boldsymbol{\Theta}},
\end{equation}
where, $\tilde{\bf E}_a\,\!^\mu$ and $\tilde{\bf E}^{(\Omega)}$ would represent, respectively, the tetrad field equations 4-form and a constraint 4-form on the conformal factor $\Omega$. $\tilde{\boldsymbol\Theta}$ represents the symplectic potential 3-form of the conformal frame after requiring that ${\delta\bf L}=0$ for any variation of the fields. The tetrad field equations $\tilde{\bf E}_a\,\!^\mu=0$ read,
\begin{align}\label{ConfE}
    \partial_\nu(\tilde{e}\tilde{\mathcal{S}}_a\,\!^{\mu\nu})&=8\pi\tilde{e}\tilde{\mathcal{J}}_a\,\!^{\mu}-\tfrac{2}{3}\Omega^2\tilde{e}\left(\tilde{\Sigma}_\nu\,\tilde{\mathcal{S}}_a\,\!^{\mu\nu}+\tilde{\Sigma}_a\tilde{T}^\mu\right)\nonumber\\
    &\quad+\Omega^2\tilde{e}\left(\tilde{\Sigma}_a\,\!^{\lambda\nu}\tilde{T}^\mu\,\!_{\nu\lambda}-2\tilde{\Sigma}_a\,\!^{\mu\nu}\tilde{T}_\nu\right)\nonumber\\
    &\quad-\tfrac{2}{3}\Omega^4\tilde{e}\left(\tilde{\Sigma}_a\tilde{\Sigma}^\mu-\tfrac{1}{2}\tilde{e}^\mu_a\tilde{\Sigma}_\nu\tilde{\Sigma}^\nu\right)\nonumber\\
    &\quad+2\Omega^2\partial_\nu\left(\tilde{e}\;\!\tilde{\Sigma}_a\,\!^{\mu\nu}\right),
\end{align}
For convenience, we have introduced here the ``induced" torsion tensor, $\tilde{\Sigma}^a\,\!_{\mu\nu}=\Omega^{-3}\left(e^a_\mu\tilde\nabla_\nu\Omega-e^a_\nu\tilde\nabla_\mu\Omega\right)$. The trace of the latter is given by $\tilde{\Sigma}_\mu\equiv\tilde{\Sigma}^\nu\,\!_{\mu\nu}=-3\Omega^{-3}\tilde\nabla_\mu\Omega$. 

The symplectic potential then reads,
\begin{align}\label{ConfSymplectic}
    \tilde{\boldsymbol\Theta}_{\beta\gamma\lambda}&=-\frac{1}{8\pi\Omega^2}\delta\tilde{e}^a_\mu\left(\tilde{\mathcal{S}}_a\,\!^{\mu\alpha}-2\Omega^2\tilde{\Sigma}_a\,\!^{\mu\alpha}\right)\tilde{\boldsymbol\epsilon}_{\alpha\beta\gamma\lambda}\nonumber\\
    &\quad+\frac{1}{4\pi\Omega^3}\delta\Omega\left(\tilde{T}^{\alpha}+\Omega^2\tilde{\Sigma}^\alpha\right)\tilde{\boldsymbol\epsilon}_{\alpha\beta\gamma\lambda}.
\end{align}
With these expressions at hand, we can now extract the Noether charge by building the 3-form current $\tilde{{\bf J}}=\tilde{\boldsymbol{\Theta}}-i_{\tilde{\xi}}\tilde{{\bf L}}$, from which the 2-from charge, $\tilde{\bf Q}_{\gamma\lambda}$, is easily read off:
\begin{align}\label{ConfCharge}
    \tilde{\bf Q}_{\gamma\lambda}=\frac{1}{16\pi}\frac{\tilde{\xi}^a}{\Omega^2}\left({\tilde{\mathcal{S}}_a}\,\!^{\alpha\beta}-2\Omega^2\tilde{\Sigma}_a\,\!^{\alpha\beta}\right)\tilde{\boldsymbol{\epsilon}}_{\alpha\beta\gamma\lambda}.
\end{align}
This is the Noether charge used in Sec.~\ref{sec:UnderWeyl} to examine the black hole entropy in the conformal frame.

\end{document}